\documentstyle[aps,pra]{revtex}
\begin{document}
\draft

\newcommand{\pp}[1]{\phantom{#1}}
\newcommand{\be}{\begin{eqnarray}}
\newcommand{\ee}{\end{eqnarray}}
\newcommand{\ve}{\varepsilon}
\newcommand{\vs}{\varsigma}
\newcommand{\Tr}{{\,\rm Tr\,}}
\newcommand{\pol}{\frac{1}{2}}

\title{
Darboux-integrable nonlinear Liouville-von
Neumann equation 
}
\author{Sergei~B.~Leble$^1$ and Marek~Czachor$^2$}
\address{
Katedra Fizyki Teoretycznej i Metod Matematycznych\\
 Politechnika Gda\'{n}ska,
ul. Narutowicza 11/12, 80-952 Gda\'{n}sk, Poland\\
$^1$ e-mail: leble@mifgate.pg.gda.pl\\
$^2$ e-mail: mczachor@sunrise.pg.gda.pl
}
\maketitle

\begin{abstract}
A new form of a binary Darboux transformation is used to
generate analytical solutions of a nonlinear Liouville-von Neumann equation.
General theory is illustrated by explicit examples.
\end{abstract}

\section{Introduction}

Nonlinear operator equations one encounters in 
quantum optics and quantum field theory are 
typically solved by techniques which are either perturbative
or semiclassical (cf. \cite{rev1,rev2}). 
The situation is caused by the fact that
analytic methods of dealing with ``non-Abelian" nonlinearites
are still at a rather preliminary stage of development. An
important step towards more efficient analytical techniques is
associated with the notion of an inverse spectral
transformation. The use of the method in the contex of matrix
equations can be found in \cite{1,1a,2} where an analytical treatment
of Maxwell-Bloch equations is given. 
In application to the Maxwell-Bloch system describing
three-level atoms interacting with light \cite{3} one
makes use of a degenerate Zakharov-Shabat spectral problem with
reduction constraints \cite{4}. The same problem is used
in the context of the complex modified Korteweg-de Vries (MKdV)
equation for a slowly varying envelope of electromagnetic field
in an optical fiber \cite{5}. 

A technical complication occurs if a solution obtained by an
inverse method should additionally satisfy some constraint. 
For example, it is often
essential to guarantee that the solution one gets is
Hermitian or positive. Difficultes of this kind were 
one of the motivations for
the development of new Darboux-type operator techniques of
solving non-Abelian equations. Particularly useful turned out to
be the method of elementary and binary Darboux transformations
introduced by one of us \cite{LU93,LZ97,L98}. These particular
versions of the Darboux transformations are more primitive than
the ordinary ones \cite{MS} in the sense that the latter can be
obtained by their composition. The binary transformation, 
a result of an application of two mutually
conjugated elementary Darboux transformations one after another, 
was successfully
applied to a three-state Maxwell-Bloch system with degeneracy 
in \cite{LU93}, and various multisoliton
solutions, including the well known $2\pi$-pulse and
breathers, were found.

In this article we apply a generalization of this technique \cite{L98}
to a new type of nonlinear operator equation. The nonlinear Liouville-von
Neumann equation we will discuss is the simplest nontrivial example of a
Lie-Nambu dynamics of a density matrix and occurs naturally in 
certain version of nonlinear quantum mechanics. 
To begin with, let us recall that the well
known Liouville-von Neumann equation (LvNE)
\be
i\dot\rho &=& [H,\rho],\label{LvN}
\ee
where $H$ is a Hamiltonian operator, $\rho$ a density matrix, and the
dot denotes the time derivative, is linear. In
Hartree-type theories one considers more general, nonlinear
equations of the form
\be
i\dot\rho &=& [H(\rho),\rho],\label{HLvN}
\ee
where $H(\rho)$ is a nonlinear Hamiltonian operator.
For time-independent Hamiltonians
$H\big(\rho(t)\big)=H\big(\rho(0)\big)$  the formal
solutions are exponential 
\be
\rho(t) &=& \exp\big[-iH\big(\rho(0)\big)t\big]\rho(0)
\exp\big[iH\big(\rho(0)\big)t\big].\label{sol1}
\ee
Both kinds of nonlinear LvNE's can be written in either Lie-Poisson
\cite{Bona,Jordan} or Lie-Nambu forms
\cite{Mor,MCpla,MCMM,MCijtp,MCpra98}. The Lie-Nambu version
involves a 3-bracket and the LvNE's can be written as 
\begin{eqnarray}
i \dot\rho_a=\{\rho_a,H_f,S\}=\{\rho_a,H_f\}\label{1}
\end{eqnarray}
where $\{\cdot,\cdot\}:=\{\cdot,\cdot,S\}$ is a Lie-Poisson bracket.
Here $\rho_a:=\rho_{AA'}(\bbox a,\bbox a')$ are components of
$\rho$ in some basis, $A$ and $A'$ are discrete (say, spinor)
indices and $\bbox a$, $\bbox a'$ the continuous ones. 
$H_f$ is a Hamiltonian function and $S$ a functional that can be
identified 
with the 2-entropy of Dar\'oczy \cite{Dar} and
Tsallis \cite{Tsallis}, i.e. $2S={\rm
Tr}(\rho^2)=\parallel \rho\parallel^2$ is the Hilbert-Schmidt
squared norm of $\rho$. 

An extension from a Lie-Poisson bracket to a 3-bracket led Nambu
to a generalization of classical Hamiltonian dynamics
\cite{Nambu}. The 3-bracket equation (\ref{1}) naturally leads
to the question of possible Nambu-type extensions of the
Lie-Poisson dynamics of density matrices. An interesting class of such
generalizations occurs if one keeps $H_f(\rho)$ linear in $\rho$
but $S$ is a function of other Dar\'oczy-Tsallis entropies.
Such Nambu-type equations are rather unusual from the point of
view of generalized Nambu-Poisson theories
\cite{N1,N2,N3,N3a,N3b,N3c,N4,N5,N6,N7,N8}. 
The peculiarity is that although the 3-bracket
itself does not satisfy the so-called fundamental identity,
typically regarded as a Nambu analogue of the Jacobi identity, 
the 2-bracket defined via 
$\{\cdot,\cdot\}_{H_f}:=\{\cdot,H_f,\cdot\}$ does satisfy 
the ordinary Jacobi identity if $H_f(\rho)$ is a linear
functional of $\rho$ \cite{MCijtp}. 
It follows that restricting $H_f$ to linear functionals one
effectively uses the Nambu-type structure as an intermediate
step which allows one to introduce a new Poisson structure,
and now $S$ plays a role of a Hamiltonian function. The standard
way of introducing interactions, i.e. by modifying $H_f$, looks
from this perspective as a modification of the Poisson structure
while keeping the Hamiltonian $S$ fixed. A simultaneous 
change of the two Nambu-type
generators, $H_f$ and $S$, can be regarded as a change of the
Hamiltonian function $S$, accompanied by a 
modification of the Poisson structure defined in terms of $H_f$.
In this respect the Nambu-type bi-Hamiltonian dynamics with
linear $H_f$ has a
logical structure analogous to this of general relativity. 
The choice of linear
$H_f$ and generalized 
$S$ can be also motivated by difficulties with probability
interpretation of generalized observables since there is no
{\it physically\/} natural definition of spectrum of nonlinear operators
\cite{MCpla,MCpra96,Mielnik98}. The 3-bracket structure can be
shown to be a a particular case of a still more general
$(2k+1)$-bracket one that, for $2k+1>3$, always vanishes on
pure states and therefore is invisible at the level of the
Schr\"odinger dynamics \cite{MCijtp}. 

The nonlinear LvNE corresponding to $S={\rm Tr}(\rho^n)/n$,
\be
i\dot\rho=[H,\rho^{n-1}],\label{n-eq}
\ee
was introduced in \cite{MCpla}. General properties of such
equations were discussed in \cite{MCMM} and \cite{MCijtp}. It was shown, in
particular, that spectra of their Hermitian Hilbert-Schmidt
solutions are time-independent. This opens a possibility of a
density matrix interpretation of the solutions. Let us note that
for $\rho^2=\rho$ (pure states) the equations reduce to the
linear LvNE and, therefore, the pure state dynamics is
indistinguishable from the ordinary linear Schr\"odinger one.
One of the problems that still remained open was
how to solve such nonlinear equations. There exist formal
solutions given in a form of a series, but the question of
convergence of such a series was not investigated. 

The aim of this paper is to describe an algebraic method that
leads to solutions of a nonlinear LvNE which reduces to
(\ref{n-eq}) with $n=3$ (3-entropy equation) in special cases.
The equation we shall study is 
\be
i\dot\rho=[H,\rho^2] +i\rho' H+iH\rho'\label{3-eq}
\ee
where the prime denotes a derivative with respect to  some additional
parameter $\tau$.  
We will generate the solutions from a Lax pair with the help of
a binary Darboux transformation. 
To avoid technicalities we will generally assume that the Hamiltonian $H$
and other operators are finite-dimensional matrices, but the
transformation works in a much more general setting (see the
example of the harmonic oscillator) and its
application to general infinite-dimensional systems is a subject of
current study. 

\section{Lax pair and its Darboux covariance}

The technique of Darboux-type transformations is perhaps the
most powerful analytical method of solving differential
equations. Although it was developed mainly in the context of nonlinear
equations, it is implicitly used also in standard
textbook quantum mechanics under the name of the
creation-annihilation operator method. The method of creation
operators is simultaneously a good illustration of the way the
Darboux technique works. In short, to use the method one has to begin
with an initial
solution which is found by other means (a ``ground
state"). Then one has to find a ``creation operator" and the
Darboux transformation is a systematic procedure that allows one
to do it. In linear
cases once we have these two elements, we are able to generate
an entire Hilbert space of solutions. In nonlinear cases the
spaces of solutions are bigger and therefore a given ``ground
state" and a ``creation operator" may generate only a subset of this
space. It is mainly for this reason that much effort was devoted to
finding different generalizations of Darboux transformations
(cf. \cite{MS}). The method we will use was devised for non-commutative
equations such as Heisenberg equations of motion. The
construction given in \cite{LZ97,L98} led to a transformation
more general than the one we use and its derivation from
elementary transformations is somewhat tedious. However, once one
has our explicit form, one can check by a straightforward
calculation that the binary transformation indeed maps one
solution into another. To make
this paper self-contained we give
the explicit proof in the Appendix.

Consider the following pair of Zakharov-Shabat equations
\be
i\varphi(\mu)' &=& (U-\mu H)\varphi(\mu)
=:
Z_\mu\varphi(\mu)\label{1a}\\
i\dot\varphi(\mu) &=& (UH+HU-\mu H^2)\varphi(\mu)\label{1b}\\
&=&
{\textstyle\frac{1}{\mu}}\big(U^2-Z^2_\mu\big)\varphi(\mu)
\ee
where $U$ and $H$ are Hermitian matrices the dot and prime
denote, respectively, derivatives with respect to  time $t$ and some
auxiliary parameter $\tau$, and $\mu$ is
complex. The solution $\varphi(\mu)$ is also in general a matrix. 
We assume that $H$ is $t$ and $\tau$-independent and $U=U(t,\tau)$. The
compatibility condition for (\ref{1a}), (\ref{1b}) is 
\be
i\dot U=[H,U^2]+iU'H+iHU',\label{equation}
\ee
and therefore the above pair is the Lax pair for (\ref{3-eq}).
We will stick to the notation with $U$ instead of $\rho$ since
non-Hermitian and non-positive solutions are also of some
interest and $\rho$ will be reserved for density matrices. 

We will need two additional conjugated problems 
\be
-i\psi(\lambda)' &=& \psi(\lambda)(U-\lambda H)\label{2a}\\
-i\dot\psi(\lambda) &=& \psi(\lambda)(UH+HU-\lambda H^2)\label{2b}
\ee
\be
-i\chi(\nu)' &=& \chi(\nu)(U-\nu H)\label{3a}\\
-i\dot\chi(\nu) &=& \chi(\nu)(UH+HU-\nu H^2)\label{3b}
\ee
each of them playing a role of a Lax pair for (\ref{3-eq}).

Consider for the moment the following general Zakharov-Shabat
problems 
\be
i\partial\varphi(\mu) &=& (V-\mu J)\varphi(\mu)\label{1c}\\
-i\partial\psi(\lambda) &=& \psi(\lambda)(V-\lambda J)\label{2c}\\
-i\partial\chi(\nu) &=& \chi(\nu)(V-\nu J)\label{3c}
\ee
where $\partial$ denotes a derivative with respect to some parameter.
We will take
the binary transformation in the form
\be
{}&{}&\psi[1](\lambda,\mu,\nu)\nonumber\\
&{}&\pp =
=
\psi(\lambda)\Big[
\bbox 1
-
\frac{\nu-\mu}{\lambda-\mu}
\varphi(\mu)
\Big(p\chi(\nu)\varphi(\mu)p
\Big)^{-1}
\chi(\nu)\Big]\label{bin}\\
&{}&\pp =
=:
\psi(\lambda)\Big[
\bbox 1
-
\frac{\nu-\mu}{\lambda-\mu}P\Big]\label{bin'}
\ee
where $p$ is a constant projector ($\partial p=0$)
and the inverse means an inverse in the $p$-invariant subspace:
$(pxp)^{-1}pxp=pxp(pxp)^{-1}=p$. The operator $P$ defined by
(\ref{bin'}) is idempotent ($P^2=P$) but in general non-Hermitian.
$P$ satisfies the nonlinear master equation \cite{master}
\be
i\partial P 
&=&
(V-\mu J)P-P(V-\nu J)
+
(\mu-\nu)PJP\label{master}
\ee
The binary transformation implies the following transformation
of the potential 
\be
V[1](\mu,\nu)
&=&
V+
(\mu-\nu)
[P,J].\label{V[1]}
\ee
Applying this general result to $V=U$, $J=H$ we get
\be
U[1](\mu,\nu)
&=&
U+
(\mu-\nu)
[P,H].
\label{binU}
\ee
The second triple of equations we have started with corresponds
to $V=UH+HU$ and $J=H^2$. In this case 
\be
V[1](\mu,\nu)
&=&
U[1](\mu,\nu)H+HU[1](\mu,\nu).
\ee
This means that (\ref{binU}) guarantees simultaneous
covariance of the Lax pairs under the
binary transformation (\ref{bin}). 

Another important feature of the binary transformation is the
fact that for $\nu=\bar \mu$ and $p\chi(\bar \mu)=p\varphi(\mu)^{\dag}$ the 
Hermiticity of the potential is Darboux-covariant, i.e 
\be
U[1](\mu,\bar \mu)^{\dag}=U[1](\mu,\bar \mu)
\ee
if $U^{\dag}=U$.

Using (\ref{master}) one can show by a straightforward
calculation (see the Appendix) that the binary transformed 
\be
\psi[1]=\psi\Big(\bbox 1-\frac{\bar\mu - \mu}{\lambda
-\mu}P\Big) 
\ee
indeed satisfies 
\be
-i\psi[1]' &=& \psi[1](U[1]-\lambda H)\\
-i\dot\psi[1] &=& \psi[1](U[1]H+HU[1]-\lambda H^2)
\ee
with $U[1]=U[1](\mu,\bar \mu)$ and, therefore, 
\be
i\dot U[1]=[H,U[1]^2]+iU[1]'H+iHU[1]'.
\ee
Subsequent iterations of the Darboux transformation generate
further solutions. Starting with a Hermitian solution $U$ we
obtain an infinite sequence of Hermitian solutions $U[1]$,
$U[2]$,... satisfying Tr~$U=$Tr~$U[1]=$Tr~$U[2]$...

\section{Covariance of the constraint $U'H+HU'=0$}

In order to generate solutions of $i\dot U=[H,U^2]$ one has to
maintain the constraint
$U'H+HU'=U[1]'H+HU[1]'=U[2]'H+HU[2]'=\dots=0$. 
Starting with stationary solutions 
\be
i\varphi(\mu)'=z \varphi(\mu)
\ee
one finds that $U'=0$ implies $U[1]'=0$. An alternative approach can be
applied to Hamiltonians of the Dirac type
\be
H=\bbox p\cdot\bbox \alpha +m\beta,
\ee
which satisfy $H^2=E(p)^2\bbox 1$ and therefore imply
\be
U[1]'H+HU[1]'&=&
U'H+HU'+(\mu-\bar\mu)[P',H^2]\nonumber\\
&=& 
U'H+HU'
\ee
which makes the constraint hereditary. 

In general, using (\ref{master}), one finds that the constraint
is hereditary if 
\be
[P_\perp(U-\mu H)P-P(U-\nu H)P_\perp,H^2]=0,
\ee
where $P_\perp=\bbox 1-P$. 

\section{Particular cases}

In this section we shall discuss properties of solutions
corresponding to several choices of the initial $U$.

For some applications one can restrict the general form (\ref{bin})
by choosing 
\be
p=
\left(
\begin{array}{ccc}
1 & 0\dots & 0 \\
0 & 0\dots & 0\\
\vdots & \vdots & \vdots\\
0 & 0\dots & 0
\end{array}
\right),
\quad
\varphi(\mu)=
\left(
\begin{array}{ccc}
\varphi_1 & 0\dots & 0 \\
\varphi_2 & 0\dots & 0\\
\vdots & \vdots & \vdots\\
\varphi_n & 0\dots & 0
\end{array}
\right),\label{matrixform}
\ee
$\nu=\bar\mu$ and $\chi(\bar \mu)=\varphi(\mu)^{\dag}$. 
Denoting the first column in $\varphi(\mu)$ by $|\varphi\rangle$
one finds that 
\be
U[1](\mu,\bar \mu)=U+(\mu-\bar\mu)[P,H]
\ee
where
\be
P=\frac{|\varphi\rangle\langle\varphi|}{\langle\varphi|\varphi\rangle}.
\label{proj}
\ee
The transition from the column solution $|\varphi\rangle$ to 
(\ref{matrixform}) is a useful trick that allows one to consider
expressions such as $p\varphi p$ which otherwise would not make
any sense.

\subsection{$U^2=U$, $U'=0$}

This is an interesting case since $U=U(t)$ is a solution of the
ordinary linear LvNE:
\be
U(t)=
\exp\big[-iHt\big]U(0)
\exp\big[iHt\big].\label{37}
\ee
Let us take a solution stationary with respect to $\tau$:
\be
i|\varphi(\mu)\rangle' &=& (U-\mu H)|\varphi(\mu)\rangle
=z_\mu|\varphi(\mu) \rangle
\ee
and define
\be
|\tilde\varphi\rangle=\exp\big[iHt\big]|\varphi\rangle.
\ee
The Lax pair is now
\be
z_\mu |\tilde\varphi\rangle&=& (U(0)-\mu H)|\tilde\varphi\rangle
\\
i|\dot{\tilde\varphi}\rangle
&=&
{\textstyle\frac{1}{\mu}}\big(z_\mu-z^2_\mu\big)|\tilde\varphi\rangle
\ee
with the solution 
\be
|\varphi(t,\tau)\rangle=e^{-i(Ht+\alpha_{t,\tau})}|\varphi(0,0)\rangle
\ee
where $\alpha_{t,\tau}=\frac{1}{\mu}z_\mu(1-z_\mu)t+z_\mu\tau$. 
The projector (\ref{proj}) is $\tau$-independent and satifies
the {\it linear\/} LvNE. This implies that $U[1]$ satisfies
the same linear equation as $U$. The following lemmas explain
the origin of this effect. Consider the general $P$ defined by
(\ref{bin'}) and $V[1]=V[1](\mu,\nu)$. 

\medskip\noindent
{\bf Lemma~1.}
$\partial P=0$ implies
\be
V[1]^2
&=&
V^2+(\mu-\nu)\Big(P(JV+VJ-\nu J^2)P_\perp\nonumber\\
&\pp =&
\pp{V^2+(\mu-\nu)\big(}
-P_\perp(JV+VJ-\mu J^2)P\Big).\label{L1}
\ee
{\it Proof\/}:
(\ref{master}) implies 
\be
P(V-\nu J)-(V-\mu J)P
=
(\mu-\nu)PJP
\ee
and
\be
{}&{}&(\mu-\nu)(PJPJ+JPJP-JPJ)\nonumber\\
&{}&\pp ==[P,V]J+J[P,V]+\mu J^2P-\nu PJ^2.
\ee
The latter formula leads directly to (\ref{L1}).$\Box$

\medskip\noindent
{\bf Lemma~2.} 
Assume $P'=0$ and $\dot P$ is given by (\ref{master}) with
$V=HU+UH$, $J=H^2$. Then 
\be
U[1]^2=U^2-(\mu-\nu)i\dot P.
\ee
{\it Proof\/}: 
\be
-iP_\perp\dot P 
&=&
-P_\perp(UH+HU-\mu H^2)P\\
-i\dot P P_\perp
&=&
P(UH+HU-\nu H^2)P_\perp
\ee
and therefore 
\be
U[1]^2
&=&
U^2-i(\mu-\nu)(P_\perp\dot P +\dot PP_\perp)\nonumber\\
&=&
U^2-i(\mu-\nu)\dot P.\nonumber
\ee
$\Box$

An immediate consequence of Lemma~2 is

\medskip\noindent
{\bf Lemma~3.} Assume $P'=0$ and $U^2=U$. Then 
$U[1]^2=U[1]$ if and only if $i\dot P=[H,P]$ i.e. $P$ satisfies
the linear LvNE.


\subsection{Case $U^2-aU\neq {\rm const}\cdot${\bf 1},
$[U^2-aU,H]=0$, $U'=0$} 

Let $a$ be a real number. $[U^2-aU,H]=0$ implies 
\be
U(t)=
\exp\big[-iaHt\big]U(0)
\exp\big[iaHt\big].\label{37'}
\ee
Repeating the steps from the
previous subsection we obtain
the Lax pair
\be
z_\mu |\tilde\varphi\rangle&=& (U(0)-\mu H)|\tilde\varphi\rangle
\\
i|\dot{\tilde\varphi}\rangle
&=&
{\textstyle\frac{1}{\mu}}\big(\Delta_a+az_\mu-z^2_\mu\big)
|\tilde\varphi\rangle
\ee
where $\Delta_a=U(0)^2-aU(0)$. The projector $P$ is
$\tau$-independent but possesses a nontrivial $t$-dependence
which follows from the fact that $\mu-\bar\mu\neq 0$. 
Define the function 
\be
F_a(t)=
\langle\varphi(0,0)|
\exp\Big(
i
\frac{\mu-\bar\mu}{|\mu|^2}\Delta_a t
\Big)|\varphi(0,0)\rangle\label{F}
\ee
satisfying
\be
\langle\varphi(t,\tau)|\varphi(t,\tau)\rangle=
\exp[i(\bar \alpha_{t,\tau}-\alpha_{t,\tau})]F_a(t),
\ee
where $\alpha_{t,\tau}=\frac{1}{\mu}z_\mu(a-z_\mu)t+z_\mu\tau$.
We find finally 
\be
U[1](t)
&=&
e^{-iaHt}\Big(U(0)+(\mu-\bar\mu)F_a(t)^{-1}\\
&\pp =&
\times e^{-\frac{i}{\mu}\Delta_a t}
\big[|\varphi(0,0)\rangle\langle\varphi(0,0)|,H\big]
e^{\frac{i}{\bar \mu}\Delta_a t}\Big)e^{iaHt}\nonumber\\
&=:&
e^{-iaHt}U_{\rm int}(t)
e^{iaHt}.\label{int}
\ee
Let us note that what makes $U[1](t)$ nontrivial is essentially 
the presence of
$F(t)$ in the denominator. It is precisely this property of the binary
Darboux transformation that is responsible for the soliton 
solutions in the Maxwell-Bloch case \cite{LU93}.

\section{Examples}

We shall now demonstrate on explicit examples how the method
works. We will concentrate on the first Darboux transformation
$U[1]$. Further iterations, $U[2]$, $\dots$, $U[n]$, are also
interesting and their relation to $U[1]$ is similar to this
between solitons and multi-solitons. The problem will be
discussed in a forthcoming paper. 

\subsection{$3\times 3$ matrix Hamiltonian, $a=1$}

Consider the Hamiltonian
\be 
H=
\left(
\begin{array}{ccc}
0 & 1 & 0\\
1 & 0 & 0\\
0 & 0 & \frac{1}{\sqrt{2}}
\end{array}
\right)\label{H}
\ee
and take $\mu=i$ (for real $\mu$ the binary transformation is
trivial). We begin with 
\be 
U(0)=
\left(
\begin{array}{ccc}
\frac{1}{2}+ \frac{\sqrt{2}}{2} & 0  & 0\\
0 & \frac{1}{2}- \frac{\sqrt{2}}{2}  & 0\\
0 & 0 & \frac{1}{2}
\end{array}
\right).
\ee
$U(0)$ does not commute with $H$ but 
\be 
U(0)^2-U(0)=U(t)^2-U(t)=
\frac{1}{4}
\left(
\begin{array}{ccc}
1 & 0  & 0\\
0 & 1  & 0\\
0 & 0 & -1
\end{array}
\right)
\ee
does. The eigenvalues of $U(0)-iH$ are $z_\pm=(1\pm i\sqrt{2})/2$
and $z_-$ has degeneracy 2. The two orthonormal eigenvectors
corresponding to $z_-$ are
\be
|\varphi_1\rangle
=
\left(
\begin{array}{c}
0\\
0\\
1
\end{array}
\right),
\quad
|\varphi_2\rangle
=
\frac{1}{\sqrt{2}}
\left(
\begin{array}{c}
e^{i\pi/4}\\
1\\
0
\end{array}
\right).
\ee
Taking 
\be
|\varphi(0,0)\rangle=
\frac{1}{\sqrt{2}}\Big(|\varphi_1\rangle+|\varphi_2\rangle\Big)
\ee
we get $F(t)=\cosh(t/2)$ and the internal part defined by
(\ref{int}) is given explicitly by
\be
{}&{}&
U_{\rm int}(t)\nonumber\\
&{}&\pp =
=
\left(
\begin{array}{ccc}
\frac{1 + \sqrt{2}}{2} - \frac{\sqrt{2}}{1 + e^t} & 0 &
   \frac{-1 - i}{2\sqrt{2}\cosh(t/2)}\\
0 & \frac{1 - \sqrt{2}}{2} + \frac{\sqrt{2}}{1 + e^t} & 
\frac{1}{2\cosh(t/2)}\\ 
\frac{-1 + i}{2\sqrt{2}\cosh(t/2)} & \frac{1}{2\cosh(t/2)} & \frac{1}{2}
\end{array}
\right).
\label{int'}
\ee
One can check by an explicit calculation that (\ref{int}) with
(\ref{H}) and (\ref{int'}) is a Hermitian solution of  
\be
i\dot U[1]=[H,U[1]^2]. \label{62}
\ee
Let us note that the solution (\ref{int'}) corresponds to an
initial condition $U[1](0)$ which is different from $U(0)$ and
is no longer block-diagonal in the basis block-diagonalizing
$H$. 
This is a consequence of the fact that $P$ is not block
diagonal, a fact that explains the importance of the degeneracy
condition for $z_-$ (had we chosen $z_+$ we would have obtained a
$(2\times 2)\oplus 1$ block-diagonal $P$).
The eigenvalues of $U[1](t)$ are nevertheless the same as those
of $U(0)$. This follows immediately from the $t$-independence of
spectrum of $U[1](t)$ and the fact that $U[1](t)$ tends
asymptotically to $U(t)$ for $t\to +\infty$. 
As a consequence $U[1](t)$ is neither normalized (Tr$U[1]\neq 1$)
nor positive and hence cannot be regarded as a density matrix.
It is, however, very easy to obtain a density matrix solution
once we know $U[1](t)$. The problem reduces to generating
a new solution whose spectrum is shifted
with respect to the original one by a number. This can be
accomplished by a gauge transformation. Indeed,
\be
\tilde U[1]=
e^{-2i\lambda Ht}\Big(U[1]+\lambda\bbox 1\Big)
e^{2i\lambda Ht}
\ee
is also a solution of (\ref{62}), and its spectrum is shifted by
$\lambda$ with respect to this of $U[1]$. Such positive
solutions can be regarded as non-normalized density matrices and
are sufficient for a well defined probability interpretation of
the theory. 
Let us finally note that the fact that spectrum of a
Hermitian solution is conserved by the dynamics is not
accidental but follows from general
properties of Lie-Nambu equations \cite{MCMM}.

\subsection{$3\times3$ Hamiltonian with equally-spaced spectrum}

Consider the Hamiltonian ($k$, $m\in\bbox R$)
\be 
H=
\left(
\begin{array}{ccc}
k+m & -m & 0\\
-m & k+m & 0\\
0 & 0 & k+m
\end{array}
\right),\label{Hm}
\ee
whose eigenvalues are $k$, $k+m$, $k+2m$, 
and take $\mu=i$. We begin with a non-normalized density matrix
\be 
\rho(0)=
\left(
\begin{array}{ccc}
\frac{1}{2}(a+\sqrt{4b+a^2}) & 0  & 0\\
0 & \frac{1}{2}(a-\sqrt{4b+a^2})  & 0\\
0 & 0 & c
\end{array}
\right),
\ee
satisfying
\be 
\rho(0)^2-a\rho(0)=\rho(t)^2-a\rho(t)=
\left(
\begin{array}{ccc}
b & 0  & 0\\
0 & b  & 0\\
0 & 0 & c(c-a)
\end{array}
\right).
\ee
Eigenvalues of $\rho(0)-iH$ are $z_0=c-i(k+m)$, $z_\pm=
\frac{1}{2}\big(a\pm \sqrt{a^2+4(b-m^2)}\big)-i(k+m)$. We need
this spectrum to satisfy a degeneracy condition: $z_0=z_+$ or 
$z_0=z_-$ with $c$ real and non-negative. Positivity of $\rho(0)$
requires also that $a>0$, $a-\sqrt{4b+a^2}\geq 0$. We will 
require that $b\neq 0$ (otherwise we will not get a nontrivial
$\rho[1]$) so that the parameters finally satisfy 
$0<4m^2<a^2+4b<a^2$. Let us note that $c(c-a)=b-m^2$
independently of the choice of sign in the degeneracy condition
$z_0=z_\pm$.  

Denote by $|k+m\rangle$ the joint eigenstate of $H$ (with
eigenvalue $k+m$) and $\rho(0)-iH$ (with eigenvalue $c-i(k+m)$);
the corresponding projector is $P_{k+m}=|k+m\rangle\langle
k+m|$. Let $\bbox 1_3=|k\rangle\langle
k|+|k+m\rangle\langle
k+m|+|k+2m\rangle\langle
k+2m|$, where the three projectors project on eigenstates of
$H$. We can write
\be
\Delta_a=b \bbox 1_3-m^2 P_{k+m}.
\ee
The two eigenstates corresponding to the degenerate eigenvalue 
$c-i(k+m)$ are othogonal. One of them is simply
$|\varphi_1\rangle=|k+m\rangle$; the other one is
$|\varphi_2\rangle=\phi_k|k\rangle+\phi_{k+2m}|k+2m\rangle$,
where the explicit form of $\phi_j$ is for the moment
irrelevant. 
Now take
$|\varphi(0,0)\rangle=A|\varphi_1\rangle+B|\varphi_2\rangle$, 
$|A|^2+|B|^2=1$. 
We find
\be
F_a(t)&=&e^{-2bt}\Big(1+(e^{2m^2t}-1)|A|^2\Big).
\ee
The above formulas can be also written as 
\be
H&=&\sum_{n=0,m,2m} (k+n)|k+n\rangle\langle k+n|\\
\rho(0)&=&\frac{a}{2}\Big(|k\rangle\langle k|+
|k+2m\rangle\langle k+2m|\Big)
+
c|k+m\rangle\langle k+m|\nonumber\\
&\pp =&
-\frac{1}{2}\sqrt{4b+a^2}
\Big(|k+2m\rangle\langle k|+|k\rangle\langle k+2m|\Big).
\ee
The solution is 
\be
\rho[1](t)
&=&
\rho(t)+2im \Big(1+(e^{2m^2t}-1)|A|^2\Big)^{-1}\nonumber\\
&\pp =&\times\Big(
|B|^2(\bar \phi_{k+2m}+\bar \phi_k)(\phi_{k+2m}-\phi_k)
|k,t\rangle\langle k+2m,t|\nonumber\\
&\pp =&+
\frac{1}{\sqrt{2}}e^{m^2t}
A\bar B(\bar \phi_{k+2m}+\bar \phi_k)
|k,t\rangle\langle k+m,t|\nonumber\\
&\pp =&+
\frac{1}{\sqrt{2}}e^{m^2t}
A\bar B(\bar \phi_{k+2m}-\bar \phi_k)
|k+2m,t\rangle\langle k+m,t|\nonumber\\
&\pp =&-
{\rm H.c.}\Big)
\ee
where $|k+j,t\rangle=e^{-iajt}|k+j\rangle$.

\subsection{1-dimentional harmonic oscillator}

We begin with the Hamiltonian
\be
H=\sum_{n=0}^\infty \hbar\omega \big(
{\textstyle\frac{1}{2}}+n\big)
|{\textstyle\frac{1}{2}}+n\rangle\langle {\textstyle\frac{1}{2}}+n|.
\ee
One can directly apply the construction from the above example. 
We have to choose some three-dimensional subspace which defines
$\rho(0)$. Put $k=\frac{1}{2}+l$ ($l,\,m\in \bbox N$), and
$\mu=i/(\hbar\omega )$.  The solution is 
\be
\rho[1](t)
&=&
\rho(t)+2im \Big(1+(e^{2\omega m^2t}-1)|A|^2\Big)^{-1}\nonumber\\
&\pp =&\times\Big(
|B|^2(\bar \phi_{k+2m}+\bar \phi_k)(\phi_{k+2m}-\phi_k)
|k,t\rangle\langle k+2m,t|\nonumber\\
&\pp =&+
\frac{1}{\sqrt{2}}e^{\omega m^2t}
A\bar B(\bar \phi_{k+2m}+\bar \phi_k)
|k,t\rangle\langle k+m,t|\nonumber\\
&\pp =&+
\frac{1}{\sqrt{2}}e^{\omega m^2t}
A\bar B(\bar \phi_{k+2m}-\bar \phi_k)
|k+2m,t\rangle\langle k+m,t|\nonumber\\
&\pp =&-
{\rm H.c.}\Big)
\ee
$\rho[1]$ has interesting asymptotic properties. Assume $A\neq 0$.
For $t\gg 0$ $\rho[1](t)\approx \rho(t)$ which suggests that the
nonlinear effect is transient. However, for $t\ll 0$ 
\be
\rho[1](t)
&\approx&
\rho(t)+2im 
\Big(
(\bar \phi_{k+2m}+\bar \phi_k)(\phi_{k+2m}-\phi_k)
|k,t\rangle\langle k+2m,t|
-
{\rm H.c.}\Big).
\ee
It follows that the asymptotic dynamics of $\rho[1](t)$ is linear
but around $t=0$ some sort of ``phase transition" occurs, and
the result of this transition is stable. Let us also note that
the linear evolution is determined by $\exp (-ia Ht)$ with
$|a|>2m$ and $m\in \bbox N$ . The choice of $a$ is related to the initial
condition. We obtain, therefore, an effective nonlinear
modification of frequency of the oscillator. 

Let us finally make $\phi_j$ explicit. Assume $l=0$, $m=1$,
$a=5$, $b=-4$, $z_0=z_+$ (i.e. $c=(5+\sqrt{5})/2)$, $A=B=1/\sqrt{2}$. Now
\be
\rho(0)&=&\frac{5}{2}\Big(|{\textstyle\frac{1}{2}}\rangle\langle
{\textstyle\frac{1}{2}}|+ 
|{\textstyle\frac{5}{2}}\rangle\langle {\textstyle\frac{5}{2}}|\Big)
+
\frac{5+\sqrt{5}}{2}|{\textstyle\frac{3}{2}}\rangle\langle
{\textstyle\frac{3}{2}}|
-\frac{3}{2}
\Big(|{\textstyle\frac{5}{2}}\rangle\langle
{\textstyle\frac{1}{2}}|+|{\textstyle\frac{1}{2}}\rangle\langle
{\textstyle\frac{5}{2}}|\Big) \\
|\varphi_1\rangle &=& |{\textstyle\frac{3}{2}}\rangle\\
|\varphi_2\rangle &=& -i\sqrt{\frac{3+\sqrt{5}}{6}}
|{\textstyle\frac{1}{2}}\rangle
+
\sqrt{\frac{2}{9+3\sqrt{5}}}
|{\textstyle\frac{5}{2}}\rangle.
\ee
${\rm Tr\,}\rho(0)=(15+\sqrt{5})/2$ and the eigenvalues of $\rho(0)$ are 
4, 1, and $(5+\sqrt{5})/2$. 

\subsection{Linear equation with nonlinear perturbation}

Assume $i \dot \rho_\epsilon=\epsilon [H,\rho_\epsilon^2]$ and define 
\be
\rho=\exp[-iHt]\rho_\epsilon \exp[iHt].
\ee
Then 
\be
i\dot \rho=[H,\rho] +\epsilon[H,\rho^2].
\ee
This is a Nambu-type equation obtained by taking a linear
Hamiltonian function $H_f={\rm Tr\,}(H\rho)$ and $S={\rm
Tr\,}(\rho^2)/2+ \epsilon{\rm Tr\,}(\rho^3)/3$. Average energy is, by
definition $\langle H\rangle={\rm Tr\,}(H\rho)/{\rm Tr\,}\rho$. 

Returning to the example of the harmonic oscillator we proceed
as before but now we choose $\mu=i/(\epsilon\hbar\omega)$. 
The solution becomes 
\be
\rho[1](t)
&=&
e^{-i(1+a\epsilon)Ht}\Bigg[
\rho(0)+2im \Big(1+(e^{2\epsilon\omega m^2t}-1)|A|^2\Big)^{-1}\nonumber\\
&\pp =&\times\Big(
|B|^2(\bar \phi_{k+2m}+\bar \phi_k)(\phi_{k+2m}-\phi_k)
|k\rangle\langle k+2m|\nonumber\\
&\pp =&+
\frac{1}{\sqrt{2}}e^{\epsilon\omega m^2t}
A\bar B(\bar \phi_{k+2m}+\bar \phi_k)
|k\rangle\langle k+m|\nonumber\\
&\pp =&+
\frac{1}{\sqrt{2}}e^{\epsilon\omega m^2t}
A\bar B(\bar \phi_{k+2m}-\bar \phi_k)
|k+2m\rangle\langle k+m|\nonumber\\
&\pp =&-
{\rm H.c.}\Big)\Bigg]e^{i(1+a\epsilon)Ht}.
\ee
The asymptotic dynamics is again linear and the frequency shift
is $\Delta\omega=a\epsilon\omega$. Let us note that according to
the definition of $\langle H\rangle$ the eigenvalues of energy
should be assumed to take values 
$\hbar\omega(1/2+n)$ and not $(1+a\epsilon)\hbar\omega(1/2+n)$. 

This point is essential for the probability interpretation of
such a nonlinear theory. 

\subsection{Homogeneous modification of the equation}

The equation we have solved is non-homogeneous which implies
that $\rho\mapsto {\rm const}\cdot \rho$ is not a symmetry
transformation. This fact makes it necessary to work with
non-normalized density matrices. In order to obtain a
homogeneous equation one can utilize the fact that 
${\rm Tr\,}(\rho^n)$ is time-independent (as a Casimir invariant).
Define $C(\rho)=[{\rm Tr\,}\rho/{\rm Tr\,}(\rho^3)]^{1/2}$ and consider 
\be
i\dot \rho=C(\rho)[H,\rho^2].
\ee
The equation is 1-homogeneous in $\rho$ and its solutions can be
obtained by the substitution $t\mapsto C(\rho)t$ in the
corresponding formulas given above. The multiplication of $\rho$ by
constants is a symmetry operation so that we can easily produce
solutions satisfying ${\rm Tr\,}\rho=1$. To get the equation from
the Nambu-type formalism one takes $S(\rho)=\frac{2}{3}[{\rm
Tr\,}\rho{\rm Tr\,}(\rho^3)] ^{1/2}$. 

\subsection{Two spin-1/2 particles}

The above Nambu-type formalism implies that spectra of Hermitian
solutions are time-independent. In particular, assuming that the
nonlinear dynamics is defined for a two-particle system, the
corresponding two-particle density matrix has time-independent
eigenvalues. When it comes to {\it reduced\/} density matrices
of the one-particle subsystems the situation is less simple.
Assume the two-particle system is described by the Hamiltonian 
\be
H=H_1\otimes \bbox 1+\bbox 1\otimes H_2.
\ee
On
the one hand it is clear that traces of the reduced density
matrices are time-independent. On the other hand, it
can be shown \cite{MCijtp} that 
\be
i\frac{d}{dt}\Tr_1\Big( (\Tr_2\rho)^2\Big)
&=&
2\Tr_1\Big([\Tr_2( \rho^2),\Tr_2( \rho)]  H_1\Big),\label{"BB"}
\ee
where $\Tr_k$, $k=1,\,2$ are partial traces. For $\rho^2\neq \rho$ the RHS
of (\ref{"BB"}) dos not in general vanish and this means that
the eigenvalues of the reduced density matrix $\Tr_2\rho$ can be
time-dependent. What is interesting the average energies of the
subsystems do not change as both $\Tr H_1\otimes \bbox 1\rho$
and $\Tr \bbox 1\otimes H_2\rho$ are separately conserved. It
follows that although the two subsystems do not
exchange average energy, they nevertheless exhibit some kind of
collective behavior. Since it is difficult to investigate the
effect from a general perspective, it may be instructive to
consider an explicit example of a two-particle system whose
density matrix can be explicitly calculated by the Darboux
technique. 

Consider two spin-1/2 particles described by the Hamiltonian 
\be
H=\bbox\sigma\cdot\bbox a\otimes \bbox 1+\bbox 1\otimes
\bbox\sigma\cdot\bbox b.
\ee
To make the example concrete assume that $|\bbox b|=1$ and 
$|\bbox a|=2$. We will start with the non-normalized density matrix 
\be
\rho(0)=
\frac{1}{2}
\left(
\begin{array}{cccc}
5+\sqrt{7} & 0 & 0 & 0\\
0 & 5-\sqrt{7} & 0 & 0\\
0 & 0 & 5+\sqrt{15} & 0 \\
0 & 0 & 0 & 5-\sqrt{15}
\end{array}
\right)
\ee
which is written in such a basis that 
\be
H=
2\sigma_x \otimes \bbox 1+\bbox 1\otimes\sigma_z=
\left(
\begin{array}{cccc}
1 & 2 & 0 & 0\\
2 & 1 & 0 & 0\\
0 & 0 & -1 & 2 \\
0 & 0 & 2 & -1
\end{array}
\right).
\ee
Take $a=5$. We find 
\be
\Delta_5=\rho(0)^2-5 \rho(0)
=
-\frac{1}{2}
\left(
\begin{array}{cccc}
9 & 0 & 0 & 0\\
0 & 9 & 0 & 0\\
0 & 0 & 5 & 0 \\
0 & 0 & 0 & 5
\end{array}
\right)
\ee
so that $[\Delta_5,H]=0$. Taking $\mu=i$ we find that 
$\rho(0)-iH$ has eigenvalues
$z_1=(1+i)/2$, $z_2=(1+3i)/2$, $z_3=(1-5i)/2$, where $z_1$ has
degeneracy 2. The two eigenvectors corresponding to $z_1$ are 
\be
|\varphi_1\rangle=
\frac{1}{4\sqrt{2}}
\left(
\begin{array}{c}
0 \\
0 \\
1+i\sqrt{15} \\
4
\end{array}
\right),
\quad
|\varphi_2\rangle=
\frac{1}{4\sqrt{2}}
\left(
\begin{array}{c}
-3+i\sqrt{7} \\
4\\
0\\
0
\end{array}
\right).
\ee
Assuming
\be
|\varphi(0,0)\rangle=\frac{1}{\sqrt{2}}\Big(
|\varphi_1\rangle+|\varphi_2\rangle\Big).
\ee
we obtain
\be
F_5(t)=
\frac{1}{2}\big(e^{5t}+e^{9t}\big)
\ee
and $\rho[1](t)=\exp[-5iHt]\rho_{\rm int}(t)\exp[5iHt]$ where
\be
\rho_{\rm int}(t)=
\frac{1}{2}
\left(
\begin{array}{cccc}
5-\sqrt{7}\tanh 2t & 
0&\frac{-13i - 3\sqrt{7} - \sqrt{15} -i \sqrt{105}}{8\cosh 2t} & 
   \frac{-7i + 3\sqrt{7} - 3\sqrt{15} + i\sqrt{105}}{8\cosh 2t}\\
0& 5+\sqrt{7}\tanh 2t & 
   \frac{15i + \sqrt{7} - \sqrt{15} - i\sqrt{105}}{8\cosh 2t}& 
   \frac{\sqrt{7} + \sqrt{15}}{2\cosh 2t}\\ 
  \frac{13i - 3\sqrt{7} - \sqrt{15} + i\sqrt{105}}{8\cosh 2t}& 
   \frac{-15i + \sqrt{7} - \sqrt{15} + i\sqrt{105}}{8\cosh 2t}& 
5+\sqrt{15} \tanh 2t & 0\\ 
\frac{7i + 3\sqrt{7} - 3\sqrt{15} - i\sqrt{105}}{8\cosh 2t}& 
   \frac{\sqrt{7} + \sqrt{15}}{2\cosh 2t}& 0& 
5-\sqrt{15} \tanh 2t
\end{array}
\right).
\ee
Eigenvalues $p_\pm(k)$, $k=1,\,2$, of (normalized) reduced density
matrices of the $k$-th subsystems are 
\be
p_\pm(1)&=&\frac{1}{2}\pm \frac{\sqrt{15}-\sqrt{7}}{20}
\tanh 2t\\
p_\pm(2)&=&\frac{1}{2}\pm \frac{\sqrt{26+2\sqrt{105}}}{40\cosh 2t}.
\ee
In order to check
that (\ref{"BB"}) is indeed satisfied one has to use
non-normalized density matrices (since the equation is
non-homogeneous) i.e. put 
$\Tr_1\Big( (\Tr_2\rho)^2\Big)=100\big(p_+(1)^2+p_-(1)^2\big)$.
Average energies of both subsystems are 0 for any $t$, which
also agrees with general theorems. 

The above collective phenomenon is typical of higher-entropy
dynamics and does not occur in Hartree-type equations
\cite{MCpra98}, a fact that explicitly shows that the Nambu-type
dynamics exhibits properties essentially different from those
discussed in the context of completely positive nonlinear
maps in \cite{MCMK98}. This interesting problem requires further
studies and is beyond the scope of nonrelativistic theory.

\section{Conclusions}

We have proposed an algebraic technique of solving a nonlinear
operator equation. 
The equation we have discussed can be regarded as a Heisenberg-picture
equation of motion for an operator $U$, since  writing it in the form
\be
i\dot U=[H,U^2]=[HU+UH,U]
\ee
one obtains a nonlinear 
Heisenberg equation with the time-dependent Hamiltonian $\tilde H(U)=-HU-UH$.
The choice of non-Hermitian $U$ (typical of the binary
transformation with $\nu\neq\bar\mu$) 
leads to non-Hermitian $\tilde
H$, a fact that may be of interest for a theory of open systems.

Restricting the initial solution $U$ to
projectors ($U^2=U$) we have shown that there exists a linear
orbit of the Darboux-transformation ($U[1]^2=U[1]$ and, hence,
$U[1]$ is a solution of the {\it linear\/} LvNE). This shows
incidentally that the binary transformation can be used to
generate solutions of the ordinary linear LvNE, a property that
may find applications in other contexts. 

Looking more closely at the origin of the simultaneous
covariance of both equations constituting the Lax pair,
one can immediately write other Lax pairs whose compatibility
conditions provide new nonlinear Darboux-integrable operator
equations. For example, taking the second equation with
$V=H^2U+HUH+UH^2$, $J=H^3$, and assuming the constraint $U'=0$
one obtains the compatibility condition 
\be
i\dot U=[H^2U+HUH+UH^2,U].
\ee
This highly non-Abelian nonlinear Liouville-von Neumann
(or Heisenberg) equation can be solved by the
binary Darboux transformation in a way similar to this described above.

\begin{acknowledgments}
Our work was supported by the KBN Grant No. 2~P03B~163~15. The
work of M.C. was financed in part by the Polish-Flemish 
Grant No. 007. 
\end{acknowledgments}

\section{Appendix: Proof of Darboux covariance}

We will show that (\ref{bin'}) satisfies 
\be
-i\partial\psi[1] &=& \psi[1](V[1]-\lambda J)
\ee
with $V[1]$ given by (\ref{V[1]}):
\be
{}&{}&
-i\partial 
\psi[1](\lambda,\mu,\nu)\nonumber\\
&{}&\pp ==
\psi(\lambda)(V-\lambda J)\Big[
\bbox 1
-
\frac{\nu-\mu}{\lambda-\mu}P\Big]
+
\frac{\nu-\mu}{\lambda-\mu}
\psi(\lambda)\nonumber\\
&{}&\pp {==}\times
\Big[
(V-\mu J)P-P(V-\nu J)
+
(\mu-\nu)PJP
\Big]\nonumber\\
&{}&\pp ==
\psi[1](\lambda,\mu,\nu)(V-\lambda J)
+
\frac{\nu-\mu}{\lambda-\mu}
\psi(\lambda)\nonumber\\
&{}&\pp {==}\times
\Big[-(V-\lambda J)P+
(V-\mu J)P-(\lambda -\nu)PJ\nonumber\\
&{}&\pp {==}
+
(\mu-\nu)PJP
\Big]\nonumber\\
&{}&\pp ==
\psi[1](\lambda,\mu,\nu)(V-\lambda J)
+
\frac{\nu-\mu}{\lambda-\mu}
\psi(\lambda)
\nonumber\\
&{}&\pp {==}\times
\Big[(\lambda -\mu) JP-(\lambda -\nu)PJ
+
(\mu-\nu)PJP
\Big]\nonumber\\
&{}&\pp ==
\psi[1](\lambda,\mu,\nu)(V[1]-\lambda J).\nonumber
\ee



\begin{references}
\bibitem{master}The master equation can be easily derived if one
keeps in mind that, by definition of the $p$-inverse,
$(pxp)^{-1}=p(pxp)^{-1}p$ and 
$$
\partial(pxp)^{-1}=-
(pxp)^{-1}\partial(pxp)(pxp)^{-1}.
$$
\end{references}
\end{document}